\documentclass[aps,prl,reprint,amsmath,amssymb,superscriptaddress,showpacs]{revtex4-2}
\usepackage{graphicx}
\usepackage{dcolumn}
\usepackage{bm}
\usepackage{upgreek}
\usepackage[colorlinks,linkcolor=blue,anchorcolor=blue,citecolor=blue,urlcolor=blue,filecolor=blue,menucolor=blue,runcolor=blue]{hyperref}

\begin{document}
	
\title{Electronic band reconstruction across the insulator-metal transition in colossal magnetoresistive EuCd$_2$P$_2$}
\author{Huali Zhang}
\thanks{These authors contributed equally to this work.}
\affiliation{Center for Correlated Matter and School of Physics, Zhejiang University, Hangzhou 310058, China}
\author{Feng Du}
\thanks{These authors contributed equally to this work.}
\affiliation{Center for Correlated Matter and School of Physics, Zhejiang University, Hangzhou 310058, China}
\author{Xiaoying Zheng}
\affiliation{Center for Correlated Matter and School of Physics, Zhejiang University, Hangzhou 310058, China}
\author{Shuaishuai Luo}
\affiliation{Center for Correlated Matter and School of Physics, Zhejiang University, Hangzhou 310058, China}
\author{Yi Wu}
\affiliation{Center for Correlated Matter and School of Physics, Zhejiang University, Hangzhou 310058, China}
\author{Hao Zheng}
\affiliation{Center for Correlated Matter and School of Physics, Zhejiang University, Hangzhou 310058, China}
\author{Shengtao Cui}
\affiliation{National Synchrotron Radiation Laboratory, University of Science and Technology of China, Hefei 230029, China}
\author{Zhe Sun}
\affiliation{National Synchrotron Radiation Laboratory, University of Science and Technology of China, Hefei 230029, China}
\author{Zhengtai Liu}
\affiliation{State Key Laboratory of Functional Materials for Informatics and Center for Excellence in Superconducting Electronics, SIMIT, Chinese Academy of Science, Shanghai, China}
\author{Dawei Shen}
\affiliation{National Synchrotron Radiation Laboratory, University of Science and Technology of China, Hefei 230029, China}
\author{Michael Smidman}
\affiliation{Center for Correlated Matter and School of Physics, Zhejiang University, Hangzhou 310058, China}
\author{Yu Song}
\affiliation{Center for Correlated Matter and School of Physics, Zhejiang University, Hangzhou 310058, China}
\author{Ming Shi}
\affiliation{Paul Scherrer Institute, Swiss Light Source, CH-5232 Villigen PSI, Switzerland}
\affiliation{Center for Correlated Matter and School of Physics, Zhejiang University, Hangzhou 310058, China}
\author{Zhicheng Zhong}
\affiliation{CAS Key Laboratory of Magnetic Materials and Devices and Zhejiang Province Key Laboratory of Magnetic Materials and Application Technology, Ningbo Institute of Materials Technology and Engineering, Chinese Academy of Science, Ningbo 315201, China}
\author{Chao Cao}
\affiliation{Center for Correlated Matter and School of Physics, Zhejiang University, Hangzhou 310058, China}
\author{Huiqiu Yuan}
\email{hqyuan@zju.edu.cn}
\affiliation{Center for Correlated Matter and School of Physics, Zhejiang University, Hangzhou 310058, China}
\affiliation{Collaborative Innovation Center of Advanced Microstructures, Nanjing University, Nanjing 210093, China}
\affiliation{State Key Laboratory of Silicon Materials, Zhejiang University, Hangzhou 310058, China}
\author{Yang Liu}
\email{yangliuphys@zju.edu.cn}
\affiliation{Center for Correlated Matter and School of Physics, Zhejiang University, Hangzhou 310058, China}
\affiliation{Collaborative Innovation Center of Advanced Microstructures, Nanjing University, Nanjing 210093, China}
\date{\today}
\addcontentsline{toc}{chapter}{Abstract}

\begin{abstract}
While colossal magnetoresistance (CMR) in Eu-based compounds is often associated with strong spin-carrier interactions, the underlying reconstruction of the electronic bands is much less understood from spectroscopic experiments. Here using angle-resolved photoemission, we directly observe an electronic band reconstruction across the insulator-metal (and magnetic) transition in the recently discovered CMR compound EuCd$_2$P$_2$. This transition is manifested by a large magnetic band splitting associated with the magnetic order, as well as unusual energy shifts of the valence bands: both the large ordered moment of Eu and carrier localization in the paramagnetic phase are crucial. Our results provide spectroscopic evidence for an electronic structure reconstruction underlying the enormous CMR observed in EuCd$_2$P$_2$, which could be important for understanding Eu-based CMR materials, as well as designing CMR materials based on large-moment rare-earth magnets.	
\end{abstract}

\maketitle

Spin-carrier interactions can have profound impact on the transport properties of solids. For example, in materials with colossal magnetoresistance (CMR), the resistance can exhibit a dramatic reduction by several orders of magnitude under an external magnetic field \cite{review1,review2,review3,SalamonRMP2001}. The canonical CMR systems are the manganites, where the intrinsic inhomogeneous states caused by strong electron correlations and associated phase competition are essential for the observed CMR \cite{dagotto2003nanoscale,Dagotto2005,Tokura2006}. CMR has also been observed in a number of Eu-based compounds \cite{Oliver1972,Shapira1974,Jiang2005,Zhang2020,yan2022,wang2022,Blawat2022,Li2021,Wang2021,jo20,Sullow2000JAP,Rosa2020,Souza2022}, where formation of magnetic polarons (MPs) is often thought to be important for the CMR \cite{Sullow2000JAP,Rosa2020,Souza2022,review4,Snow2001,das2012,Pohlit2018PRL,AleCrivillero2023}: the strong exchange coupling between the Eu moments and the charge carriers leads to the formation of ordered magnetic clusters (or MPs) that can trap the charge carriers; a percolation transition of MPs, caused by spontaneous magnetic order or external magnetic fields, can result in delocalization of charge carriers (hence the CMR).

Recently, an enormous CMR up to $10^5 \%$ has been observed in EuCd$_2$P$_2$ \cite{Wang2021}, attracting considerable interest \cite{Flebus2021prb,Eliot2022,Homes2023,Sunko2022}. Since no mixed valency or appreciable lattice distortion was observed, the CMR mechanism here is likely different from the classical CMR picture based on manganites. Magnetic fluctuations \cite{Wang2021} and a magnetic Berezinskii-Kosterlitz-Thouless transition \cite{Flebus2021prb,Eliot2022} have both been proposed to account for the giant CMR in EuCd$_2$P$_2$. On the other hand, the formation of ferromagnetic clusters (or MPs) have also been invoked to explain its origin \cite{Homes2023,Sunko2022}.

\begin{figure*}[ht]
	\includegraphics[width=1.\linewidth]{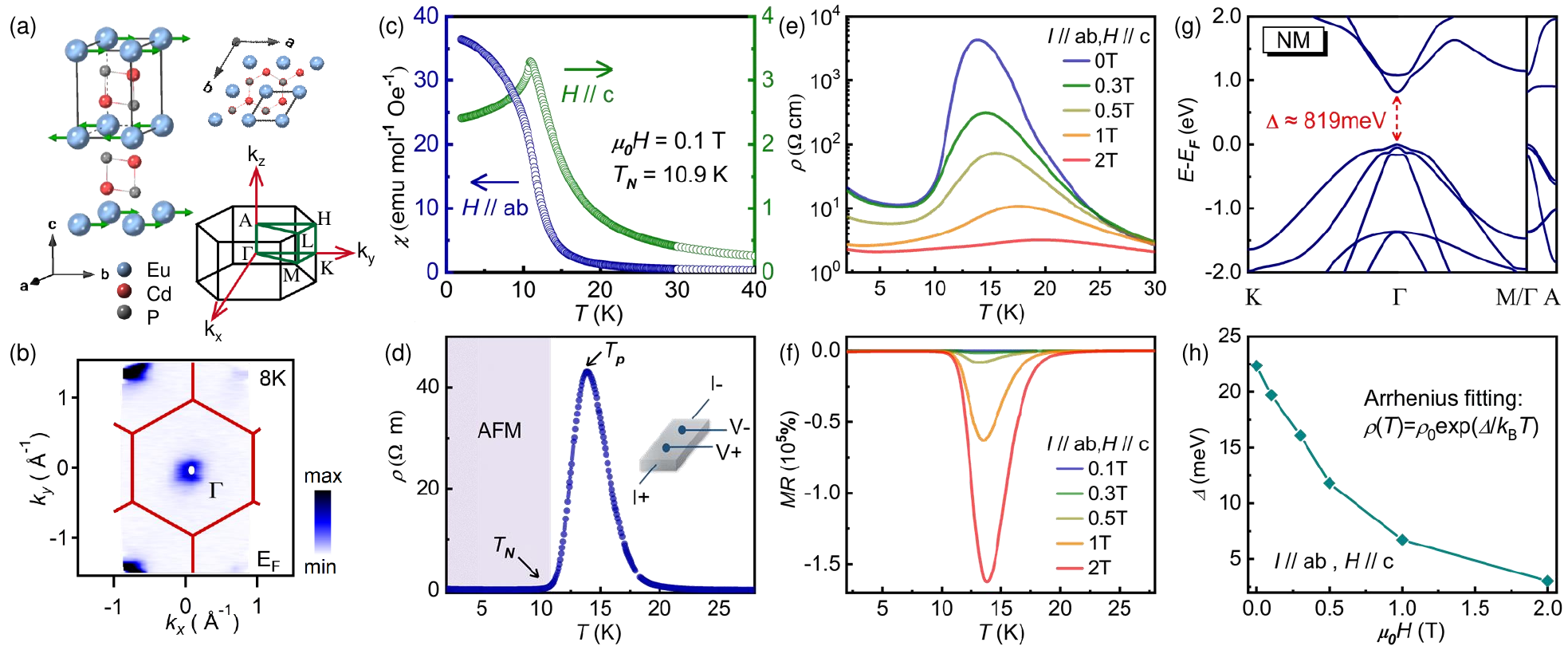}
	\centering
	\caption{(\textbf{a}) Crystal structure and Brillouin zone (BZ) of EuCd$_2$P$_2$. Crystal axes, high-symmetry momentum points and the Eu spin directions in the AFM phase are labelled accordingly. (\textbf{b}) The in-plane FS map at 8 K from ARPES measurements using 106 eV photons. The red hexagons indicate the BZ boundaries. (\textbf{c}) The magnetic susceptibility as a function of temperature for fields along the $c$ axis (green) and the $ab$ plane (blue). (\textbf{d}) The resistivity as a function of temperature. The resistivity peak at $T_p$ = 14 K is higher than $T_N$ = 10.9 K. (\textbf{e}) Temperature-dependent resistivity (log scale) at different magnetic fields. The slight upturn at very low temperature might be related to Kondo-like scattering. (\textbf{f}) MR corresponding to (\textbf{e}). (\textbf{g}) Calculated band structure along high-symmetry directions for the nonmagnetic (NM) phase. (\textbf{h}) Activation gap as a function of magnetic field, from fitting the temperature-dependent resistivity above $T_p$ in (\textbf{e}) with the Arrhenius equation. }
	\label{Fig1}
\end{figure*}

To uncover the mechanism of CMR in EuCd$_2$P$_2$, it is important to understand the origin of the sharp peak in the resistivity as a function of temperature at zero magnetic field [see Fig. \ref{Fig1}(d) and discussions below], which is often observed in CMR materials. However, in Eu-based CMR compounds, the microscopic understanding of the resistivity peak or insulator-metal transition (IMT) remains controversial. For example, in EuMn$_2$Sb$_2$ \cite{Sun2021}, EuTe$_2$ \cite{Yang2021} and pressurized EuCd$_2$As$_2$ \cite{Du2022}, a reconstruction of the band structure was proposed across the IMT which underlies the CMR, which was mainly supported by magneto-transport measurements or theoretical calculations. On the other hand, in previous photoemission studies of a similar Eu-based compound EuCd$_2$As$_2$, where the CMR is much smaller compared to EuCd$_2$P$_2$, electronic signatures associated with the magnetic order can be identified \cite{Wang2016,Ma2019,Ma2020AM,Jo2021}, but the system appears to remain metallic with a well-defined Fermi surface (FS) across the temperature range of the resistivity peak, implying that spin-dependent scattering likely plays an important role. One peculiar aspect of EuCd$_2$P$_2$ is that the temperature of the resistivity peak ($T_p$) is clearly higher than the antiferromagnetic (AFM) ordering temperature $T_N$, whereas in other Eu-based CMR materials the two temperatures often coincide. The much more pronounced resistivity peak in EuCd$_2$P$_2$ further implies stronger electronic reconstruction, allowing for direct experimental detection.

Here we present spectroscopic evidence of a distinct band structure reconstruction in EuCd$_2$P$_2$ across the IMT, using angle-resolved photoemission spectroscopy (ARPES). The details of crystal synthesis, transport measurements and density functional theory (DFT) calculations can be found in \cite{SM,PhysRevB.47.558,PhysRevB.59.1758,Yang2021NST}. EuCd$_2$P$_2$ is a layered compound with a simple hexagonal structure as shown in Fig. \ref{Fig1}(a). At high temperature, it is paramagnetic (PM) with Curie-Weiss-like behavior due to localized $4f$ moments. Below $T_N$ = 10.9 K, it exhibits long-range A-type AFM order with Eu moments pointing along the $a$/$b$ direction, i.e., the moments are aligned ferromagnetically within the $ab$ plane and have opposite directions for neighboring layers [see Fig. \ref{Fig1}(a)]. This AFM order is supported by our magnetic susceptibility data shown in Fig. \ref{Fig1}(c), where there is a peak at $T_N$ when the field is along the $c$ axis. In contrast, the magnetic susceptibility shows a ferromagnetic-like behavior for fields in the $ab$ plane (easy plane), implying that (short-range) ferromagnetic (FM) order occurs slightly above $T_N$, as verified recently by different magnetic probes \cite{Sunko2022}. The resistivity exhibits a sharp peak at $T_p$ $\sim$ 14 K [Fig. \ref{Fig1}(d)], which is obviously higher than $T_N$. Note that the absolute value of the resistivity and $T_p$ are somewhat different from those in Ref. \cite{Wang2021}, which is likely due to slight variations in the sample stoichiometry or impurity doping \cite{SM}. The strong resistivity peak at $T_p$ can be dramatically suppressed by a small magnetic field of $\sim$2 T, as shown in Fig. \ref{Fig1}(e), leading to a huge magnetoresistance as previously reported \cite{Wang2021}. Here we define the magnetoresistance (MR) as \begin{equation}MR=\frac{R_B-R_0}{R_B}\times 100\% \end{equation} and plot it as a function of temperature for different magnetic fields [Fig. \ref{Fig1}(f)]. The magnitude of the MR exceeds $10^5 \%$ with $\upmu_0$$H$ = 2 T and is much larger than other Eu-based CMR compounds, including EuIn$_2$P$_2$ \cite{Jiang2005,Zhang2020}, EuIn$_2$As$_2$ \cite{yan2022}, EuZn$_2$As$_2$ \cite{wang2022,Blawat2022}, EuSn$_2$As$_2$ \cite{Li2021} and EuCd$_2$As$_2$ \cite{Wang2021,jo20}. It might be possible to further enhance the MR in EuCd$_2$P$_2$ by careful tuning of the growth condition.

\begin{figure}[ht]
	\includegraphics[width=1.\columnwidth]{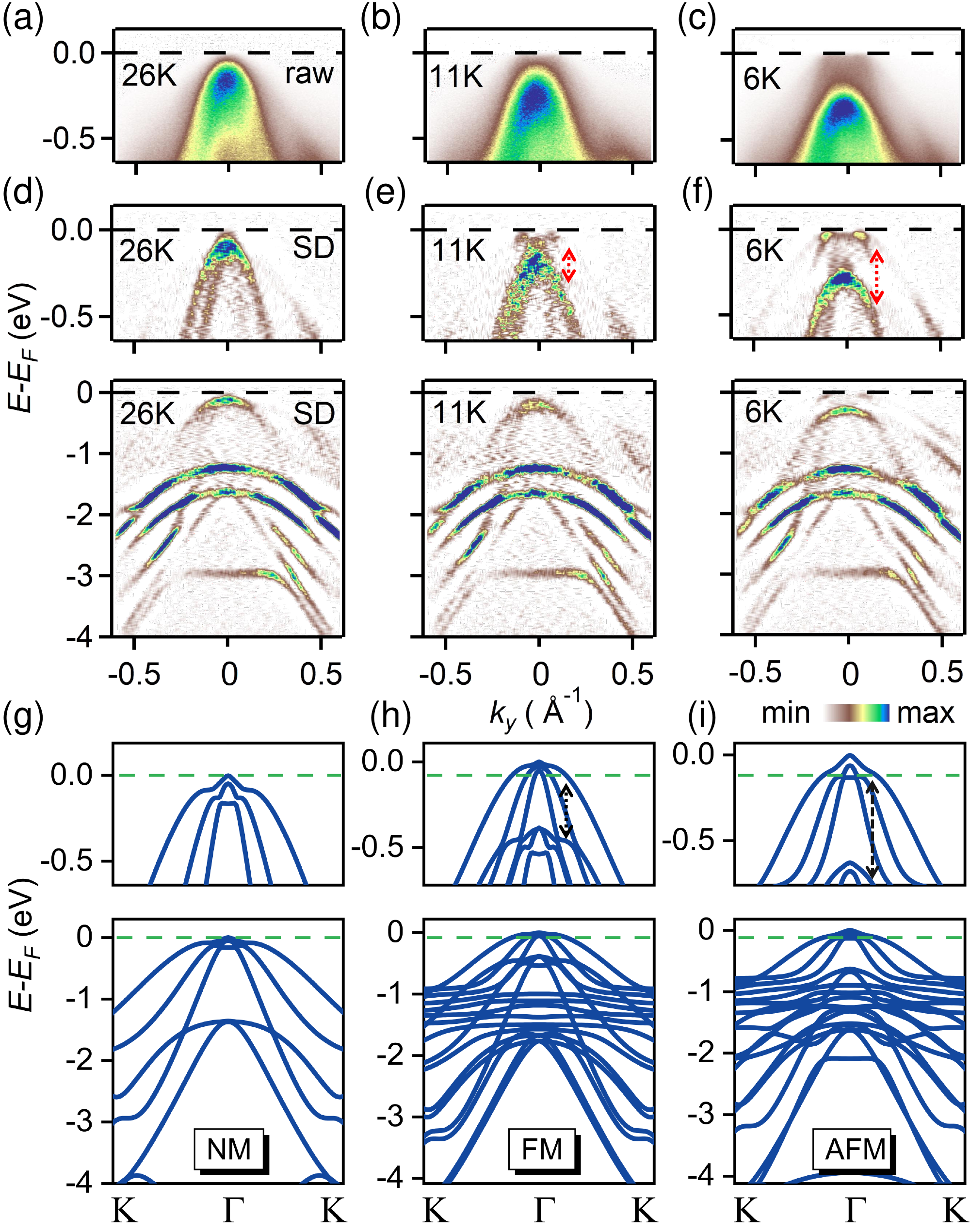}
	\centering
	\caption{(\textbf{a-c}) The ARPES data near $E_F$ along $\Gamma-K$ taken with 21.2 eV photons at three representative temperatures. (\textbf{d-f}) The corresponding second derivatives. The top and bottom rows show the data in different energy ranges. (\textbf{g-i}) Calculated band structures for the NM, FM and AFM phases, respectively, in comparison with (\textbf{d-f}). The horizontal dashed lines in (\textbf{g-i}) indicate the experimental $E_F$ positions from comparison with (\textbf{d-f}), which are 0.08 eV (FM phase) and 0.12 eV (AFM phase) lower than the calculated values. The red (black) dashed arrows in (\textbf{e,f}) [(\textbf{h,i})] indicate the experimental (calculated) band splittings compared to the NM phase. Note that the treatment of Eu $4f$ electrons are different for the NM and magnetic phases \cite{SM}. }
	\label{Fig2}
\end{figure}

Figure \ref{Fig2}(a-c) show the band structures near the BZ center at three representative temperatures from ARPES measurements, and the corresponding second derivatives are displayed in Fig. \ref{Fig2}(d-f), where the fine features can be better visualized [see Fig. S1 in \cite{SM}]. Since only the valence bands near the $\Gamma$ point contribute to the FS [see FS maps in Fig. \ref{Fig1}(b) and DFT calculations in Fig. \ref{Fig1}(g)], we focus on the spectra near $\Gamma$ obtained from 21.2 eV photons. Well above $T_p$, i.e., in the PM phase, one can identify at least two hole bands centered at the $\Gamma$ point [they can be better visualized in the second derivatives in Fig. \ref{Fig2}(d)], in reasonable agreement with the DFT calculations shown in Fig. \ref{Fig2}(g). Here the experimental valence band maximum (VBM) is slightly below $E_F$; in fact, the activation gap from the resistivity fitting using the Arrhenius equation is only $\sim$20 meV at zero field [see Fig. \ref{Fig1}(h)], which is fourty times smaller than the calculated band gap from DFT (819 meV) shown in Fig. \ref{Fig1}(g). This implies that the activation gap above $T_p$ is not between the valence and conduction bands. Instead, it arises from excitations between the valence bands and some type of localized state slightly above $E_F$ (see below).

\begin{figure*}[ht]
	\includegraphics[width=1.\linewidth]{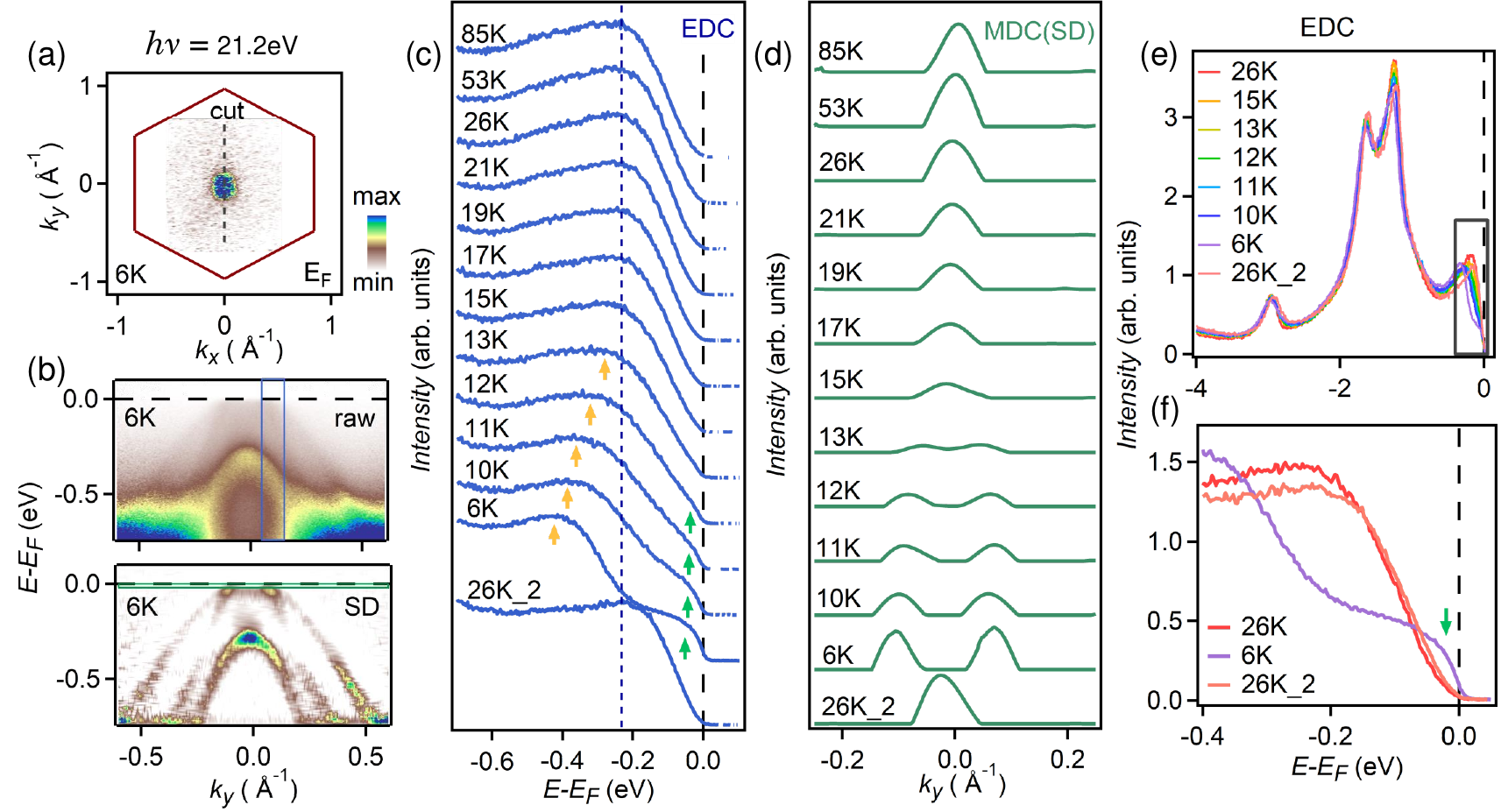}
	\centering
	\caption{ (\textbf{a}) FS map at 6 K taken with 21.2 eV photons. (\textbf{b}) The ARPES spectra (top) and the corresponding second derivative (bottom) along the cut shown in (\textbf{a}) at 6 K. (\textbf{c,d}) Temperature evolution of the EDCs (\textbf{c}) and MDCs at $E_F$ (\textbf{d}), which are offset vertically for display clarity. The EDCs in (\textbf{c}) and MDCs in (\textbf{d}) are integrated over the light blue and green rectangles shown in (\textbf{b}), respectively. Here second derivatives of MDCs are shown. (\textbf{e}) Temperature evolution of large energy-scale EDCs. (\textbf{f}) is a zoom-in view near $E_F$. 26K\_2 in (\textbf{c-f}) indicates another scan at 26 K after a cooling-warming cycle. }
	\label{Fig3}
\end{figure*}

As the temperature drops below $T_p$ (but above $T_N$), where (short-range) FM order begins to develop \cite{Sunko2022}, the valence bands begin to show signs of magnetic exchange splitting [red arrow in Fig. \ref{Fig2}(e)]: the top valence band (with weak intensity) now crosses $E_F$, while the lower valence bands (with stronger intensity) move downward and become broader due to band overlap, consistent with DFT calculations shown in Fig. \ref{Fig2}(h). Going further below $T_N$, the experimental magnetic band splitting (MBS) becomes larger [red arrow in Fig. \ref{Fig2}(f)]. Note that the calculated band splitting in the AFM phase [black arrow in Fig. \ref{Fig2}(i)], mainly due to AFM band folding, is obviously larger than the experimental value in Fig. \ref{Fig2}(f). Interestingly, the experimental splitting at 6 K is much closer to the calculated value of the FM phase [black arrow in Fig. \ref{Fig2}(h)]. Since the experimental splitting evolves smoothly from the FM phase to the AFM phase [see Fig. \ref{Fig3} for details], these experimental facts imply that the electronic states near $E_F$ are mostly sensitive to the in-plane FM order, and much less affected by the spin alignment along the out-of-plane direction.

The systematic temperature evolution is summarized in Fig. \ref{Fig3}, where Fig. \ref{Fig3}(a,b) show the associated FS and energy-momentum cut, respectively. Figure \ref{Fig3}(c) shows the temperature evolution of the energy distribution curves (EDCs): above $\sim$15 K, the EDCs do not show clear spectral weight near $E_F$, consistent with the semiconducting nature; below $\sim$15 K, the valence band at $\sim$-0.23 eV (yellow arrows) shifts to lower energies and simultaneously a small peak emerges near $E_F$ (green arrows), indicating a transition to a metallic state. Such an IMT can also be identified from the temperature evolution of the momentum distribution curves (MDCs) at $E_F$ in Fig. \ref{Fig3}(d): above $\sim$15 K, the MDC features a broad central peak from the tail of the valence band, which gradually diminishes at lower temperature due to reduced thermal population; below $\sim$15 K, the central peak splits into two peaks, whose separation becomes larger at lower temperatures, indicating the development of a well-defined FS. Here the second derivatives of the MDCs are shown for better visualization; the raw data, which show similar temperature evolution, can be found in Fig. S11 in \cite{SM}. Since a previous laser ARPES study on a relevant compound EuCd$_2$As$_2$ showed that surface absorption can obviously affect the ARPES spectra \cite{Wang2022prb}, we performed cooling-warming temperature cycles [Fig. \ref{Fig3}(c-f)], as well as time-dependent measurements [Fig. S10 in \cite{SM}], to ensure that the observed change is intrinsic and caused by temperature. Such electronic structure reconstruction is further confirmed by ARPES measurements using higher photon energies [Fig. S2-S3 in \cite{SM}], as well as measurements on samples with different resistivity curves [Fig. S9 in \cite{SM}]. The temperature evolution in Fig. \ref{Fig3} indicates that the onset of the band structure reconstruction occurs near $T_p$ (and clearly above $T_N$), and the evolution across $T_N$ is smooth without any sudden change. This is consistent with the continuous evolution of the resistivity across $T_N$, which should be caused by the electronic band reconstruction due to the in-plane FM order that precedes the AFM order \cite{Sunko2022}.

The electronic structure reconstruction across the IMT manifests through the large splittings of the valence bands, facilitated by the large ordered moments of Eu ($\sim$7 $\mu_B$/Eu), which act like a large effective magnetic field to split the valence bands. Note that the magnitude of the MBS is strongly band-dependent, e.g., for the deeper valence bands below -1 eV in Fig. \ref{Fig3}(e), the splitting is too small to be resolved experimentally. Such a band dependence can be related to the different interaction strengths with the Eu $4f$ moments, which can be reasonably captured by DFT calculations [Fig. \ref{Fig2}(h,i)]. In addition, the Eu $4f$ electrons are quite localized and do not directly contribute to the FS, since the $4f$ bands are located at $\sim$-1 eV [see Fig. S2 in \cite{SM}]. Our results also rule out the possibility of appreciable Kondo effect or valence transition (associated with Eu $4f$ electrons) as the cause of the resistivity peak.

Another key ingredient for the electronically driven IMT is the energy shift of the valence bands related to the magnetic order [see green dashed lines in Fig. \ref{Fig2}(g-i) for comparison with DFT]. According to DFT calculations, the band gap of EuCd$_2$P$_2$ is $\sim$600 meV and $\sim$400 meV for the AFM and FM phases, respectively [see Fig. S4 in \cite{SM}]. This implies that the observed temperature-dependent shift of the valence bands and their crossings of $E_F$ cannot be explained by DFT calculations, as the position of the VBM in DFT would be fixed relative to $E_F$ due to constant electron filling [Fig. \ref{Fig2}(g-i)]. One likely explanation for the band shift is the proposed MPs \cite{Sullow2000JAP,Rosa2020,Souza2022,review4,Snow2001,das2012,Pohlit2018PRL,AleCrivillero2023,Homes2023,Sunko2022}: the existence of MPs in the PM phase could potentially give rise to a gap near $E_F$ with a localized state lying slightly above the VBM, which effectively traps the charge carriers; upon development of magnetic order, the MPs become percolated, liberating trapped hole-like carriers (note that the upward shift of the valence bands implies filling of mobile hole carriers at low temperatures). Another possible explanation for the band shift is p-type impurity doping in the sample. In this case, the localized states (or acceptor levels) should shift from above $E_F$ in the PM phase to below $E_F$ in the magnetic phases \cite{Seo2021}, allowing some valence holes to be occupied. However, these acceptor levels are not directly observed in the current experiments, possibly due to their random distribution within the lattice and very low concentration.

\begin{figure}[ht]
	\includegraphics[width=1.\columnwidth]{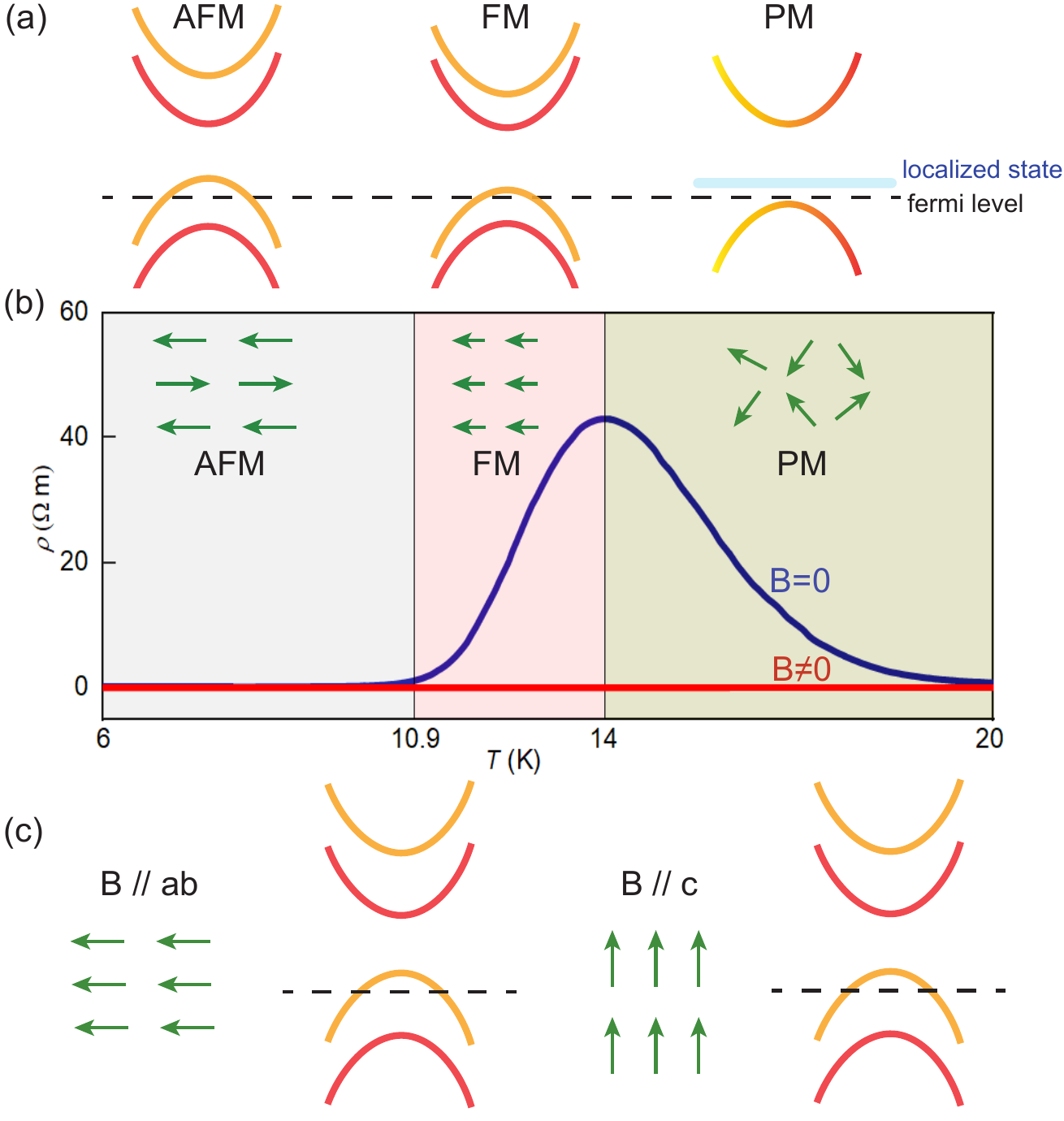}
	\centering
	\caption{(\textbf{a}) Schematic view of the experimental band structures near $E_F$ for the three phases at zero magnetic field. The light blue line indicates the localized state. Note that the experimental MBS evolves smoothly from FM to AFM and it is mainly determined by the in-plane FM order. (\textbf{b}) Temperature-dependent resistivity with and without an applied magnetic field. Three regions corresponding to (\textbf{a}) are labelled. (\textbf{c}) Schematic view of the magnetic structures and corresponding band structures in the presence of magnetic fields in the $ab$ plane (left) and along the $c$ axis (right). }
	\label{Fig4}
\end{figure}

Figure \ref{Fig4} demonstrates schematically the key findings of our study. In the PM phase, the system is semiconducting with a small activation gap, i.e., $E_F$ lies between the VBM and the localized state [Fig. \ref{Fig4}(a)]. As the (short-range) FM order sets in near $T_p$, moderate MBS takes place, leading to the IMT. Now the split valence bands cross $E_F$ and become mobile charge carriers. Below $T_N$, the band splitting increases continuously, resulting in a very metallic state with more charge carriers. The large MBS in the magnetic phases and the small activation gap in the PM phase are both important for the electronic IMT observed experimentally.

The aforementioned reconstruction of the electronic structure also provides a natural explanation for the enormous negative CMR observed in this compound [Fig. \ref{Fig4}(b,c)]. Under an external magnetic field (along the $c$ axis or within the $ab$ plane), the Eu moments will be polarized along the field directions [see Fig. S5 in \cite{SM}], leading to a field-induced FM state. DFT calculations indeed confirm that the MBS is of similar magnitude for Eu moments pointing along the $c$ or $a/b$ directions [see Fig. S4 in \cite{SM}]. Therefore, a metallic FM ground state is expected under a large external magnetic field, consistent with the vanishing activation gap in transport [Fig. \ref{Fig1}(h)].

It is interesting to compare EuCd$_2$P$_2$ with other similar Eu-based materials, e.g., EuCd$_2$As$_2$ \cite{Ma2019,Ma2020AM,Jo2021}, where MBS was also observed although no IMT in the band structure has been identified. It would be desirable to check whether a tiny energy gap opens near $E_F$ in the PM phase in these materials. In addition, the pronounced CMR in EuCd$_2$P$_2$ is facilitated by an activation gap of appropriate size in the PM state, although the detailed mechanism of charge localization in the PM phase, particularly the role of MPs, deserves further studies.

To conclude, using high-resolution ARPES, we observe a clear reconstruction of the electronic band structure underlying the sharp resistivity peak in colossal magnetoresistive EuCd$_2$P$_2$. Our results demonstrate that the resistivity peak is caused by a large MBS mainly from the in-plane FM order, as well as associated shift of the valence bands. Both the large ordered moments of Eu and carrier localization in the PM phase are crucial for the observed CMR. The physics here is apparently different from the canonical scenario based on manganites, where strong electron correlation leads to intrinsic inhomogeneity and locally metallic states in the globally insulating PM phase \cite{Sun2007,Sun2008prb}. Finally, the spectroscopic insight obtained here could be important for understanding other Eu-based CMR systems, as well as designing new CMR materials based on large-moment rare-earth elements, where large MBS and exchange coupling between $4f$ moments and conduction electrons are present.

~
\par This work is supported by the National Key R$\&$D Program of China (Grant No. 2022YFA1402200), the Key R$\&$D Program of Zhejiang Province, China (2021C01002), the State Key project of Zhejiang Province (No. LZ22A040007), the National Science Foundation of China (No. 12174331) and the Fundamental Research Funds for the Central Universities (2021FZZX001-03). Part of the APRES measurements were performed at synchrotron facilities including BL13U beamline in Hefei National Synchrotron Radiation Laboratory and beamline 03U of the Shanghai Synchrotron Radiation Facility, which was supported by ME2 Project (Grant No. 11227902) from the National Natural Science Foundation of China.

\end{document}